# Breeding the Cat Through Superposition of Two Schrödinger Kittens Based on Coupled Waveguides

Nuo Wang[1], Xinchen Zhang[1], Qi Liu[1,2], Fengxiao Sun[1], Qiongyi He[1,2,3,4,5], and Ying Gu[1,2,3,4,5,*]

[1] *State Key Laboratory for Mesoscopic Physics,*
*Department of Physics, Peking University, Beijing 100871, China*
[2] *Frontiers Science Center for Nano-optoelectronics & Collaborative Innovation*
*Center of Quantum Matter & Beijing Academy of Quantum Information Sciences,*
*Peking University, Beijing 100871, China*
[3] *Collaborative Innovation Center of Extreme Optics,*
*Shanxi University, Taiyuan, Shanxi 030006, China*
[4] *Peking University Yangtze Delta Institute of Optoelectronics, Nantong 226010, China*
[5] *Hefei National Laboratory, Hefei 230088, China*
* ygu@pku.edu.cn

**Abstract**

Optical Schrodinger's cat (SC) is highly anticipated because of the potential of realizing fault-tolerant quantum computing, but the practical merit is only shown when the amplitude is larger than 2. However, such high-amplitude cats have not been prepared due to the limitations rooted in the existing method. Here, we demonstrate a principle that a large SC-like state can be generated by the superposition of two kittens in which two nearby coherent states interfere and grow to an enlarged coherent-like state. Further, we propose a scheme to breed the cat beyond the limitation in the former works with a high probability by realizing the superposition of two SCs in coupled waveguides. The principle and scheme demonstrated here provide a new perspective on understanding quantum superposition in phase space and a better solution for the efficient generation of SCs on chips.

## 1. Introduction:

1935, Erwin Schrodinger imagined the famous Gedankenexperiment [1] known as 'Schrodinger's Cat' in which the cat in a box is in both the 'alive state' and 'dead state'. Nowadays, the SC has been abstracted as the superposition of two macroscopically distinguishable quantum states [2]. Following this idea, the cat state is defined as the superposition of two coherent states with opposite amplitude in optics [2,3]. Two special instances are $SC_\pm(\alpha) = \mathcal{N}(|\alpha\rangle \pm |-\alpha\rangle)$, which refer to even and odd cats, respectively. From the SC's birth, it has been entrusted with the mission of exploring the border between the quantum and classical world [4–7]. With further research, the optical cat state, which is a highly nongaussian state, has been demonstrated to be an important quantum resource in quantum computation [8–11], quantum metrology [12–14], and quantum teleportation [15,16]. But whatever the fundamental research or the quantum application, the amplitude of the cat is required to be high in most situations. A widely accepted criterion is that the cat should be greater than 2 so that the near-orthogonality of the qubit encoded on coherent states can be guaranteed.

To this day, many different approaches for generating the cat states have been proposed and realized. In which, schemes based on optical nonlinearity [3,17,18] and cavity Quantum Electrodynamics (QED) [19] can deterministic create the cat state or entangled cat state. Unfortunately, provided currently available experiment conditions, these two methods are strongly affected by the loss, resulting in an unsatisfactory performance. Instead, the healded techniques based on linear optics and conditional measurement, though probabilistic, stand out from these methods because of the relatively simple experimental setup and good realistic behavior. Among these, photon-subtraction from a squeezed vacuum [20–23] is a well-established method of approximately generating relatively large cat states, but the probability of detecting many photons is very low. And the quantum state engineering method [24–27] can only prepare a relatively small cat-like state because controlling the high number Fock states on demand is difficult. The amplification scheme based on multi-photon quantum interference [28–30] is promising to solve these problems. However, the ability to efficiently produce a useful cat of this method is insufficient, thus even the best experiment outcome falls short of meeting the requirement of quantum computation. So, there is an urgent need for innovative approaches that can efficiently generate higher-amplitude cat states.

Here, we reveal a principle that the superposition of two SCs can yield an enlarged SC-like state. Under reasonable superposition coefficients, the interference between two nearby coherent states will produce a coherent-like state with a greater amplitude so that the superposition of two SCs will become an amplified SC-like state. Further, based on this idea, we propose a scheme for preparing large-amplitude SC in coupled waveguides. Two kittens are input, and the output is the superposition of two SCs with amplitude less than 2 under conditional measurement. At last, an SC-like state with an amplitude greater than 2 is achieved. Compared to the previous scheme [28–30], we obtain a relatively larger cat state with a high probability under the same input. The key to our work is the superposition of two cats so we can break through the theoretical amplification limit in previous approaches. The principle advances our knowledge of quantum superposition in cat states and can be extended to other macroscopic quantum states, e.g. squeezed cat states. The breeding scheme based on coupled waveguides will be helpful to the follow-up quantum information missions on chips.

## 2. The mechanism of breeding the cat by coherent superposition

We found that an SC-like state can be generated by the superposition of two cats, i.e., $\mathcal{N}[c_1(|\alpha_1\rangle \pm |-\alpha_1\rangle) + c_2(|\alpha_2\rangle \pm |-\alpha_2\rangle)] \approx \mathcal{N}'(|\alpha_3\rangle \pm |-\alpha_3\rangle)$, where $\mathcal{N}$ and $\mathcal{N}'$ are normalization factors, as shown in Fig. 1. According to the coefficients $c_1$ and $c_2$, there are three different cases of the generated state's size, including amplification and reduction.

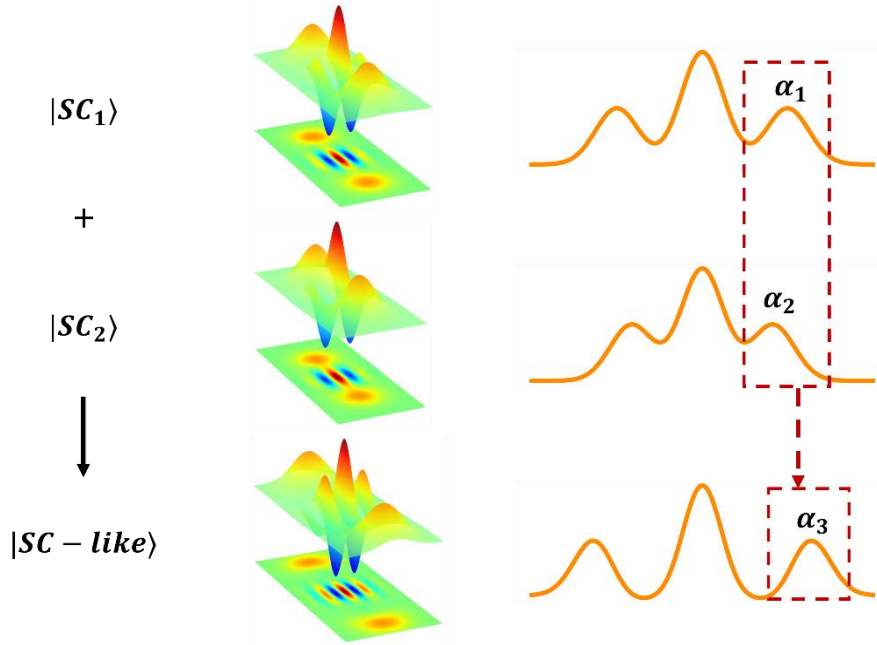

Fig. 1. Schematic of generating a SC-like state by superpositioning two SC states. The central (right) part is the three states' Wigner function (Wigner function's cross-section). The dashed boxes mean the superposition of two nearby coherent states ($|\alpha_1\rangle$ and $|\alpha_2\rangle$) will generate a coherent-like state $|\alpha_3\rangle$.

In this section, we will first illustrate the principle of generating a cat-like state by superposition. Following the principle, we can get the size relation between the three cats in different cases. Then, the condition of generating a larger SC-like by superposition can be obtained.

**2.1 The principle of generating a cat-like state by superposition**

The principle of generating an SC-like state by superposition is that two nearby coherent states' superposition will become a coherent-like state. The superposition of two SC states has been studied in [31], in which two cats superpose with equal weight, and the Wigner function of the superposed state including six interference terms has been deduced. However, the results given by these interference terms are complex, and it's hard to understand the phenomena we found intuitively in such a way. Here, we use another perspective to explain it. When the size difference between two cats is smaller than the size a lot, we can divide this phenomenon into two parts. First, as shown in the right part of Fig.1, two nearby coherent states superpose and become a coherent-like state, i.e. $\mathcal{N}(c_1|\alpha_1\rangle + c_2|\alpha_2\rangle) \approx |\alpha_3\rangle, \mathcal{N}(c_1|-\alpha_1\rangle + c_2|-\alpha_2\rangle) \approx |-\alpha_3\rangle$ where $\mathcal{N}$ will be used to represent the normalization factor uniformly in this paper. Second, two coherent-like states superpose and generate the cat-like state. From this perspective, we can understand the generation of cat-like states clearly and transform this process into a simpler one of two coherent states' superposition to analyze the condition of obtaining a larger cat. It should be noted that when $|\alpha_1 - \alpha_2|$ is very large or $c_2 \approx -c_1$ so that the Wigner function of $\mathcal{N}(c_1|\alpha_1\rangle + c_2|\alpha_2\rangle)$ has three peaks with an evident negative value, the final generated state is unable to approximate a cat state. The parameters range that we care about should not be located in these situations.

## 2.2 The size relation between the three cat states and the amplification condition

According to the above analysis, the size relation of the three cats is the same as that of the corresponding coherent states. So, in the following, we elucidate the size relation between the generated coherent-like state $|\alpha_3\rangle$ and the coherent states $|\alpha_1\rangle, |\alpha_2\rangle$ from the perspective of interference. First, the density operator of the superposed state $\mathcal{N}(c_1|\alpha_1\rangle + c_2|\alpha_2\rangle)$ can be deduced:

$$|\psi\rangle = \mathcal{N}(c_1|\alpha_1\rangle + c_2|\alpha_2\rangle),$$
$$\rho = \mathcal{N}^2(|c_1|^2|\alpha_1\rangle\langle\alpha_1| + |c_2|^2|\alpha_2\rangle\langle\alpha_2| + c_2c_1^*|\alpha_2\rangle\langle\alpha_1| + c_1c_2^*|\alpha_1\rangle\langle\alpha_2|) \quad (1)$$

For simplicity and without generality, we suppose that $\alpha_1, \alpha_2$ are real numbers and $\alpha_1 > \alpha_2 > 0$, the Wigner function can be obtained as:

$$W(\alpha) = W\big(\mathcal{N}^2(|c_1|^2|\alpha_1\rangle\langle\alpha_1| + |c_2|^2|\alpha_2\rangle\langle\alpha_2| + c_2c_1^*|\alpha_2\rangle\langle\alpha_1| + c_1c_2^*|\alpha_1\rangle\langle\alpha_2|)\big)$$
$$= \frac{2}{\pi}\mathcal{N}^2\Big\{|c_1|^2\exp(-2|\alpha-\alpha_1|^2) + |c_2|^2\exp(-2|\alpha-\alpha_2|^2)$$
$$+ 2Re(c_1c_2^*)\exp\left(-2\left|\alpha - \frac{\alpha_1+\alpha_2}{2}\right|^2\right)\cos\left[2(\alpha_1-\alpha_2)\times Im\left(\alpha - \frac{\alpha_1+\alpha_2}{2}\right)\right]\Big\} \quad (2)$$

In which, we define $W_1 = \frac{2}{\pi}\mathcal{N}^2|c_1|^2\exp(-2|\alpha-\alpha_1|^2)$, $W_2 = \frac{2}{\pi}\mathcal{N}^2|c_2|^2\exp(-2|\alpha-\alpha_2|^2)$, and $W_{int} = \frac{4}{\pi}\mathcal{N}^2 Re(c_1c_2^*)\exp\left(-2\left|\alpha - \frac{\alpha_1+\alpha_2}{2}\right|^2\right)\cos\left[2(\alpha_1-\alpha_2)\times Im\left(\alpha - \frac{\alpha_1+\alpha_2}{2}\right)\right]$. $W_1, W_2$ are corresponding to the coherent states $|\alpha_1\rangle, |\alpha_2\rangle$ respectively, and $W_{int}$ is the interference term. Here, we define $X = Re(\alpha), P = Im(\alpha)$ and use the $W(X,0)$ to analyze the size relation between $\alpha_1, \alpha_2$ and $\alpha_3$. So, $W_1(X,0) = \frac{2}{\pi}\mathcal{N}^2|c_1|^2\exp(-2|X-\alpha_1|^2)$, $W_2(X,0) = \frac{2}{\pi}\mathcal{N}^2|c_2|^2\exp(-2|X-\alpha_2|^2)$, and $W_{int}(X,0) = \frac{4}{\pi}\mathcal{N}^2 Re(c_1c_2^*)\exp\left[-2\left|X - \frac{\alpha_1+\alpha_2}{2}\right|^2\right]$. Three different cases can exist when we consider the sign of the interference term $W_{int}(X,0)$ and the proportion of two coherent states, as shown in Fig. 2:

*Case I*: $\alpha_3 > \alpha_1, \alpha_2$ when $c_1c_2 < 0, |c_1| > |c_2|$, i.e., a coherent-like state larger than $|\alpha_1\rangle, |\alpha_2\rangle$ can be generated when the superposed phase is $\pi$ and the large coherent state $|\alpha_1\rangle$ is prime in superposition. As shown in Fig 2(a), the state $|\alpha_1\rangle$ and $W_1$ are dominant when $|c_1| > |c_2|$, but the final Wigner function is a bit deviates from a Gaussian profile because of the interference. The interference term $W_{int}(X,0)$ is negative when $c_1c_2 < 0$ and located in $\frac{\alpha_1+\alpha_2}{2}$, so the destructive interference makes the peak position be "repelled" from $\alpha_1$ to a larger value.

*Case II*: $\alpha_3 < \alpha_1, \alpha_2$ when $c_1c_2 < 0, |c_1| < |c_2|$, i.e., a coherent-like state smaller than $|\alpha_1\rangle, |\alpha_2\rangle$ can be generated when the superposed phase is $\pi$ and the small coherent state $|\alpha_2\rangle$ is prime. As shown in Fig. 2(b), the state $|\alpha_2\rangle$ and $W_2$ are dominant when $|c_1| < |c_2|$ and the interference term $W_{int}(X,0)$ is negative when $c_1c_2 < 0$ and located in $\frac{\alpha_1+\alpha_2}{2}$, so the destructive interference makes the peak position

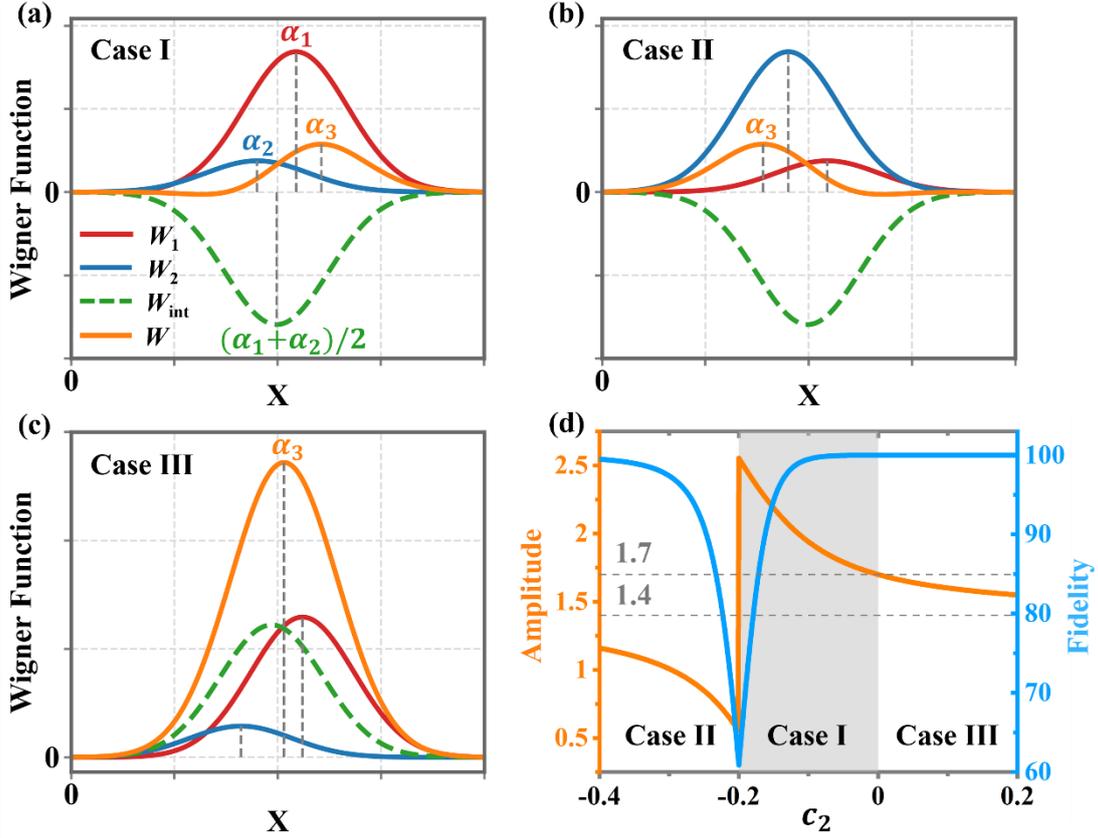

Fig. 2. The size relations between the generated coherent-like state $|\alpha_3\rangle$ and original coherent states $|\alpha_1\rangle, |\alpha_2\rangle$. The decomposition of the Wigner function (at P = 0) of the superposed state $\mathcal{N}(c_1|\alpha_1\rangle + c_2|\alpha_2\rangle)$ for different cases: (a) $c_1c_2 < 0, |c_1| > |c_2|$ (b) $c_1c_2 < 0, |c_1| < |c_2|$ and (c) $c_1c_2 > 0$. (d) The amplitude (orange curve) and fidelity (blue curve) of the coherent-like state with $c_2$ varying from $-0.4$ to $0.2$. Here, (a) - (c) are just for schematics, and (d) is a specific example with $c_1 = 0.2, \alpha_1 = 1.7, \alpha_2 = 1.4$.

be "repelled" from $\alpha_2$ to a smaller value.

*Case III*: $\alpha_2 < \alpha_3 < \alpha_1$ when $c_1c_2 > 0$, i.e., a coherent-like state between $|\alpha_1\rangle$ and $|\alpha_2\rangle$ can be generated when the superposed phase is 0. As shown in Fig. 2(c), the interference term $W_{int}(X, 0)$ is positive when $c_1c_2 > 0$ and located in $\frac{\alpha_1+\alpha_2}{2}$, so the constructive interference makes the peak position attracted to a value in between.

To enhance the universality of these results, a specific example (The correlated data can be found in Appendix A) of the amplitude and fidelity of the generated state with $\alpha_1 = 1,7, \alpha_2 = 1.4, c_1 = 0.2$, and $c_2$ caries from $-0.4$ to $0.2$ is shown in Fig. 2(d) where the three cases above are included. Moreover, the repulsive interaction is stronger but fidelity is decreased when the coefficients are closer to $c_2 = -c_1$.

In this section, we reveal the principle of generating a cat-like state by superpositioning two cats. First, two nearby coherent states overlap, interfere, and become a coherent-like state. Then, two coherent-like states superpose and generate a cat-like state. Following this perspective, by analyzing the interference of coherent states $|\alpha_1\rangle, |\alpha_2\rangle$, we can get the size relation between the three cats: when the superposed phase is $\pi$, i.e., $c_1 c_2 < 0$, the interference exhibits a repulsive interaction resulting in either $\alpha_3 > \alpha_1, \alpha_2$ ($|c_1| > |c_2|$) or $\alpha_3 < \alpha_1, \alpha_2$ ($|c_1| < |c_2|$); when the superposed phase is 0, i.e., $c_1 c_2 > 0$, the interference exhibits an attractive interaction resulting in $\alpha_2 < \alpha_3 < \alpha_1$. Finally, we can get the condition of generating an enlarged, high-fidelity, cat-like state: ① the size difference between the two cats is not huge. ② $c_1 c_2 < 0, |c_1| > |c_2|$ and not so close to $c_2 = -c_1$. Actually, the superposition of the two cats is more likely to be a squeezed cat state, which will be discussed in our subsequent research. The principle we proposed can be extended to other macroscopic quantum states, such as squeezed cat states. When it is combined with other schemes to generate cats, its potential can be further tapped, as we will demonstrate below.

## 3. Preparing large-size SC states by superpositioning two cats based on coupled waveguides

To the best of our knowledge, there are no schemes that can realize such a superposition. However, we found that the superposition of two SC states can be realized based on coupled waveguides and photon number resolving (PNR) detection. Then, a scheme based on the above principle has been proposed to break through the theoretical amplification limit in previous work [28–30]. In this section, we will first introduce the scheme of superpositioning two cat stats based on coupled waveguides. A specific example of generating a large amplitude cat state is also shown.

### 3.1 The scheme of realizing the cat state's superposition

Our scheme is based on the quantum beam splitting of coupled waveguides and PNR measurement, as shown in Fig. 3, two SC states with amplitude $\alpha_0, i\beta_0$ are input, and the superposition of two cats with amplitude $\cos(\mu z)\alpha_0 + \sin(\mu z)\beta_0, \cos(\mu z)\alpha_0 - \sin(\mu z)\beta_0$ are generated in waveguide 1 when zero photons are detected at waveguide 2. The transmission matrix of coupled waveguides is $U = \begin{bmatrix} \cos(\mu z) & -i\sin(\mu z) \\ -i\sin(\mu z) & \cos(\mu z) \end{bmatrix}$ [32], where $\mu$ is the coupling strength between two waveguides, and $z$ is the coupling length, so the effective Hamiltonian and the time evolution operator can be obtained as $\hat{H} = \hbar\mu(\hat{a}_1^\dagger \hat{a}_2 + \hat{a}_2^\dagger \hat{a}_1)$, $\hat{S}(z) = exp\left(-\frac{i\hat{H}z}{\hbar}\right) = \exp(-i\mu(\hat{a}_1^\dagger \hat{a}_2 + \hat{a}_2^\dagger \hat{a}_1)z)$, where $\hat{a}_j$ and $\hat{a}_j^\dagger$ ($j = 1,2$) are the boson annihilation and creation operators respectively. Let the initial cat states be:

$$\begin{aligned}|SC(\alpha_0)\rangle &= \mathcal{N}(|\alpha_0\rangle + e^{i\varphi_1}|-\alpha_0\rangle)\\ |SC(i\beta_0)\rangle &= \mathcal{N}(|i\beta_0\rangle + e^{i\varphi_2}|-i\beta_0\rangle)\end{aligned} \quad (3)$$

where $\alpha_0, \beta_0$ are positive real numbers, so the output state after propagating z in coupled waveguides is the superposition of four direct product states of coherent states with pure real amplitude and pure imaginary amplitude respectively:

$$\begin{aligned}\hat{S}(z)[|SC(\alpha_0)\rangle \otimes |SC(i\beta_0)\rangle] = \mathcal{N}(\; &|\cos(\mu z)\alpha_0 + \sin(\mu z)\beta_0\rangle |i[\cos(\mu z)\beta_0 - \sin(\mu z)\alpha_0]\rangle \\ +&e^{i(\varphi_1+\varphi_2)}|-[\cos(\mu z)\alpha_0 + \sin(\mu z)\beta_0]\rangle |-i[\cos(\mu z)\beta_0 - \sin(\mu z)\alpha_0]\rangle \\ +&e^{i\varphi_1}|-\cos(\mu z)\alpha_0 + \sin(\mu z)\beta_0\rangle |i[\cos(\mu z)\beta_0 + \sin(\mu z)\alpha_0]\rangle \\ +&e^{i\varphi_2}|\cos(\mu z)\alpha_0 - \sin(\mu z)\beta_0\rangle |-i[\cos(\mu z)\beta_0 + \sin(\mu z)\alpha_0]\rangle \;)\end{aligned} \quad (4)$$

where the amplitudes of the corresponding coherent state in different direct product states are opposite, which is the key to creating cat states. To get the superposition of cat states, PNR detection is done in the output port of waveguide 2, so the final output quantum state in waveguide 1:

$$\begin{aligned}|\psi_{out}(z)\rangle = \mathcal{N}\Big[\; &e^{-\frac{1}{2}(t\beta_0-r\alpha_0)^2}(|t\alpha_0+r\beta_0\rangle + e^{i(\varphi_1+\varphi_2)}|-(t\alpha_0+r\beta_0)\rangle) \\ +&e^{-\frac{1}{2}(t\beta_0+r\alpha_0)^2}(e^{i\varphi_2}|t\alpha_0-r\beta_0\rangle + e^{i\varphi_1}|-(t\alpha_0-r\beta_0)\rangle)\;\Big]\end{aligned} \quad (5)$$

where $t = \cos(\mu z), r = \sin(\mu z)$, and the detection photon number is set to zero. Measuring other photon numbers also can realize the superposition of cats (See Appendix B for detail), here, without loss of generality, we focus on the simplest situation of detecting zero photons. By the way, the phase $i$ in the input state $|SC(i\beta_0)\rangle$ is necessary for superpositioning two cats on a line in phase space. In other situations, for example, four-component cat states would be produced when the phase of the amplitude of two input cats is the same, which will be studied in our later research.

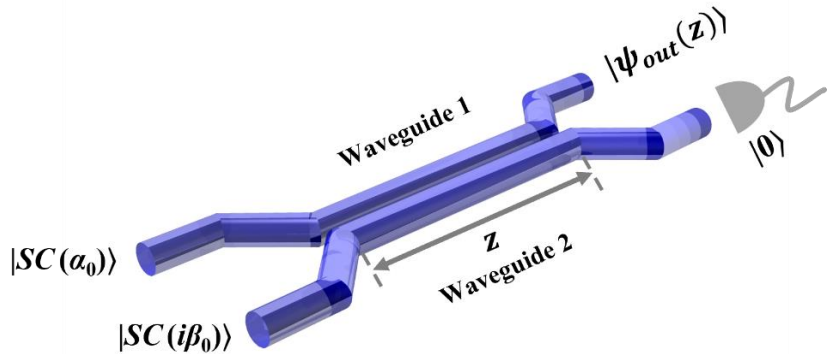

Fig. 3. The scheme of realizing the superposition of two SCs based on coupled waveguides. The generated SC-like state $|\psi_{out}(z)\rangle$ can be amplified beyond $\sqrt{\alpha_0^2 + \beta_0^2}$ under appropriate design. The coupling length is labeled as $z$ and zero photons are detected at the waveguide 2.

Here, we consider a simple but useful situation of inputting two odd cat states. Let $\varphi_1 = \varphi_2 = \pi$, so the output is the superposition of two even cat states:

$$\begin{aligned}|\psi_{out}(z)\rangle = \mathcal{N}\Big[\; &e^{-\frac{1}{2}(t\beta_0-r\alpha_0)^2}(|t\alpha_0+r\beta_0\rangle + |-(t\alpha_0+r\beta_0)\rangle) \\ -&e^{-\frac{1}{2}(t\beta_0+r\alpha_0)^2}(|t\alpha_0-r\beta_0\rangle + |-(t\alpha_0-r\beta_0)\rangle)\;\Big]\end{aligned} \quad (6)$$

It can be expressed as a familiar form:

$$|\psi_{out}(z)\rangle = \mathcal{N}[c_1(|\alpha_1\rangle + |-\alpha_1\rangle) + c_2(|\alpha_2\rangle + |-\alpha_2\rangle)]$$

$$\begin{cases} c_1 = e^{-\frac{1}{2}(t\beta_0 - r\alpha_0)^2}, c_2 = -e^{-\frac{1}{2}(t\beta_0 + r\alpha_0)^2}, \alpha_1 = |t\alpha_0 + r\beta_0|, \alpha_2 = |t\alpha_0 - r\beta_0|; r > 0 \\ c_1 = e^{-\frac{1}{2}(t\beta_0 + r\alpha_0)^2}, c_2 = -e^{-\frac{1}{2}(t\beta_0 - r\alpha_0)^2}, \alpha_1 = |t\alpha_0 - r\beta_0|, \alpha_2 = |t\alpha_0 + r\beta_0|; r < 0 \end{cases} \quad (7)$$

which satisfies that $c_1 c_2 < 0, |c_1| > |c_2|$. So, according to the mechanism in section 2, the condition of generating a larger $(\alpha_3 > \alpha_1, \alpha_2)$, high-fidelity, cat-like state by superpositioning two cats can be fulfilled here. By choosing a reasonable coupling length z, the amplitude $\alpha_3$ of the cat-like state can satisfy $\alpha_{3max} > \alpha_{1max} = \alpha_{2max} = \sqrt{\alpha_0^2 + \beta_0^2} > \alpha_0, \beta_0$. Here, $\alpha_{1max} = \alpha_{2max} = \sqrt{|\alpha_0|^2 + |i\beta_0|^2} = \sqrt{\alpha_0^2 + \beta_0^2}$ is the theoretical limit of the cat state's amplification in the previous method, which has been surpassed by our scheme.

### 3.2 A specific example: generating a cat larger than 2 with a high success probability

A specific example is given below to display the strong amplification ability of our scheme, where a cat-like state larger than 2 is generated with a high probability while such a large cat can't be achieved in previous schemes with the same input. Let the input in Fig. 3 be $\alpha_0 = 1.7, \beta_0 = 0.8$, then the amplitude, and fidelity (the fidelity between the generated states $|\psi_1\rangle$ and target state $|\psi_2\rangle$ is defined as $F = |\langle\psi_1|\psi_2\rangle|$) of the generated cat-like state, superposition coefficients, and success probability can be obtained based on the above analytical results, as shown in Fig. 4. For simplicity, let $\mu = 1$, and focusing on the range of $z = 0$ to $\pi/2$ because of periodic, the coefficients satisfy that $c_1 c_2 < 0, |c_1| > |c_2|$, so the amplitude of the SC-like state is always larger than the $|SC_1\rangle = \mathcal{N}(|t\alpha_0 + r\beta_0\rangle + |-(t\alpha_0 + r\beta_0)\rangle)$ and $SC_2 = \mathcal{N}(|t\alpha_0 - r\beta_0\rangle + |-(t\alpha_0 - r\beta_0)\rangle)$. Based on this, we can obtain a cat-like state larger than 2 in a wide range of $z$ from 0 to $0.21\pi$ while the theoretical limit in previous work with the same size input is $1.88 < 2$. The largest cat with an amplitude equal to 2.16 is obtained near $z = 0$, but the success probability is very low. By mainly considering the size and probability, we choose a relative optimal coupling length of $z = 0.14\pi$ with the amplitude of the generated state is 2.10, the success probability is 39.5% (the success probability in [28] is 29.3% with the same size input), and the fidelity is 98.31%. So, whatever the size or the success probability, our scheme is superior to the previous proposals.

It should be pointed out that although the prior schemes have considered a similar situation of generating a large cat by measurement after two kittens are inputted into a BS, they have just got a cat with a maximum size equal to $\sqrt{|\alpha_0|^2 + |i\beta_0|^2}$. The limitation in the size of these works is that they have only cared about one cat in the generated state and decreased the proportion of another state as far as possible, so the amplification ability of superposition has been omitted. Moreover, we have chosen two cats with different sizes in our example, that's because we want the interference between two cats to be strong when $|SC_1\rangle$ approaches its maximum at $z = \cos^{-1}\left(\frac{\alpha_0}{\sqrt{\alpha_0^2 + \beta_0^2}}\right)$. The interference can be characterized by the ratio of coefficients $R = \left|\frac{c_1}{c_2}\right| = e^{2tr\alpha_0\beta_0}$ which obtains its maximum value $R_{max} = e^{\alpha_0\beta_0} > 1$ at $z = \frac{\pi}{4}$.

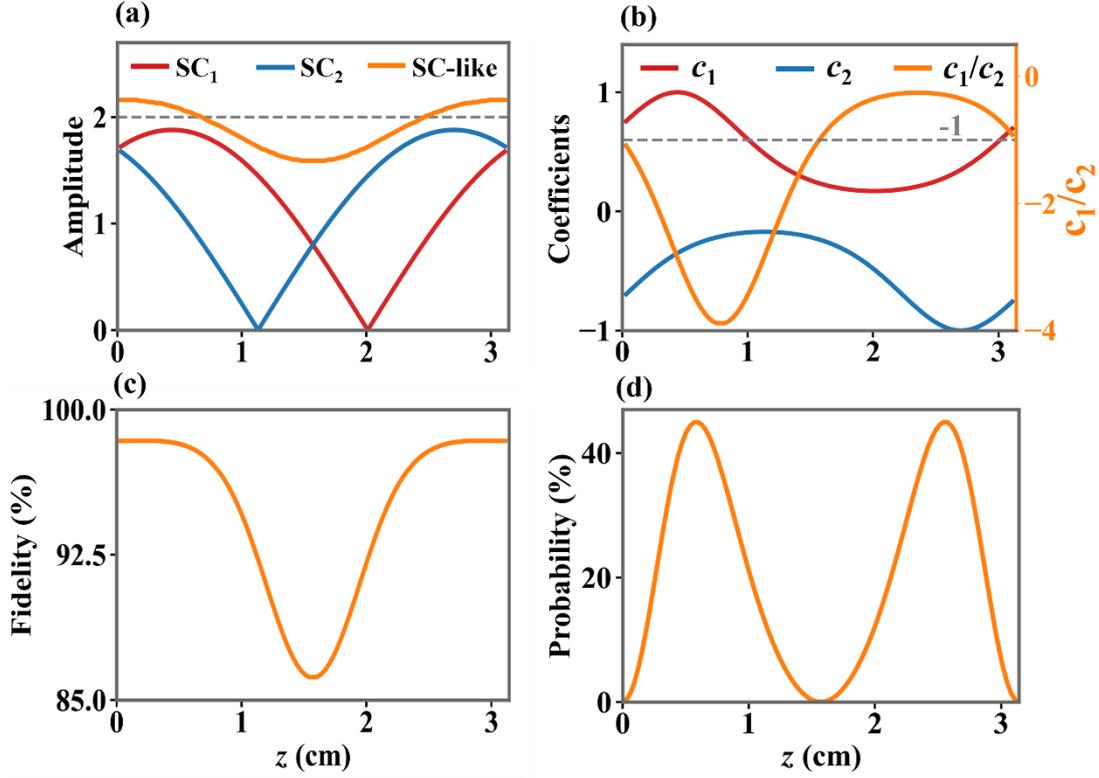

Fig. 4. The characters of SC-like state generated through the superposition of two even $|SC_1\rangle$ and $|SC_2\rangle$ in coupled waveguide system. The (a) amplitude and (b) fidelity of the generated SC-like state. (c) The superposition coefficients of $|SC_1\rangle$ and $|SC_2\rangle$ and their ratio. (d) The success probability with zero photons is detected at the output port of waveguide 2. Here, the input states are two odd SCs with $\alpha_0 = 1.7, \beta_0 = 0.8$ and zero photons are detected.

When $R$ is far from 1, the interference is small, and vice versa. So, if we want the interference to be strong when $|SC_1\rangle$ achieve a high amplitude, the position $z = \cos^{-1}\left(\frac{\alpha_0}{\sqrt{\alpha_0^2+\beta_0^2}}\right)$ should away from $z = \frac{\pi}{4}$. This can be satisfied when $\alpha_0 \neq \beta_0$, so we choose two cats with $\alpha_0 = 1.7$, and $\beta_0 = 0.8$.

## 4. Discussion: The impact of parity of the input cat states

In section 3, there are 3 different cases when considering two cat state's superposition. Correspondingly, they may exit in our coupled waveguide scheme when inputting cats with different parity. So, there will be a significant difference even bad effects. Next, we will demonstrate that our scheme performs well in most of these situations by carefully designing the input state and coupled length, which indicates good universality.

When two even cats input into the coupled waveguide, the output is the superposition of two even cats with $c_1 c_2 > 0$ which belongs to case III. So, the generated cat satisfies that $\alpha_2 < \alpha_3 < \alpha_1$, which can't be amplified effectively. Let the parameters of the input be $\varphi_1 = \varphi_2 = 0$, so according to Eq. (5), the output is:

$$|\psi_{out}(z)\rangle = \mathcal{N}\left[e^{-\frac{1}{2}(t\beta_0 - r\alpha_0)^2}(|t\alpha_0 + r\beta_0\rangle + |-(t\alpha_0 + r\beta_0)\rangle)\right.$$
$$\left. + e^{-\frac{1}{2}(t\beta_0 + r\alpha_0)^2}(|t\alpha_0 - r\beta_0\rangle + |-(t\alpha_0 - r\beta_0)\rangle)\right] \quad (8)$$

It can be expressed as:

$$|\psi_{out}(z)\rangle = \mathcal{N}[c_1(|\alpha_1\rangle + |-\alpha_1\rangle) + c_2(|\alpha_2\rangle + |-\alpha_2\rangle)]$$

$$\begin{cases} c_1 = e^{-\frac{1}{2}(t\beta_0 - r\alpha_0)^2}, c_2 = e^{-\frac{1}{2}(t\beta_0 + r\alpha_0)^2}, \alpha_1 = |t\alpha_0 + r\beta_0|, \alpha_2 = |t\alpha_0 - r\beta_0|; r > 0 \\ c_1 = e^{-\frac{1}{2}(t\beta_0 + r\alpha_0)^2}, c_2 = e^{-\frac{1}{2}(t\beta_0 - r\alpha_0)^2}, \alpha_1 = |t\alpha_0 - r\beta_0|, \alpha_2 = |t\alpha_0 + r\beta_0|; r < 0 \end{cases} \quad (9)$$

which satisfy that $c_1 c_2 > 0$, so the amplitude of the SC-like state is always between the $|SC_1\rangle = \mathcal{N}(|t\alpha_0 + r\beta_0\rangle + |-(t\alpha_0 + r\beta_0)\rangle)$ and $SC_2 = \mathcal{N}(|t\alpha_0 - r\beta_0\rangle + |-(t\alpha_0 - r\beta_0)\rangle)$. Let $\alpha_0 = 1.7, \beta_0 = 0.8$ the same as the parameters in Fig. 4, concrete results can be acquired, as shown in Fig. 5. It's obvious that the generated cat-like state is much smaller than 2.

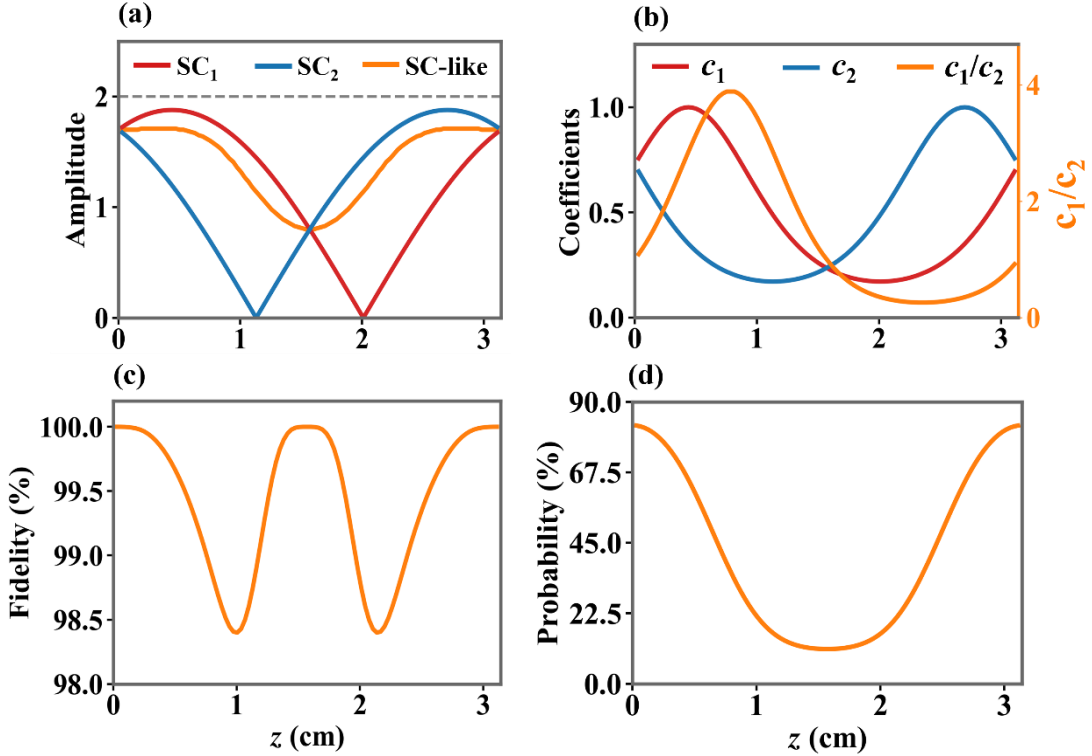

Fig. 5. The characters of SC-like state generated through the superposition of two even $|SC_1\rangle$ and $|SC_2\rangle$ in coupled waveguide system. Here, the input states are two even SCs with $\alpha_0 = 1.7, \beta_0 = 0.8$ respectively. The (a) amplitude and (b) fidelity of the generated SC-like state. (c) The superposition coefficients of $|SC_1\rangle$ and $|SC_2\rangle$ and their ratio. (d) The success probability when zero photons are detected at the output port of waveguide 2.

When an even cat is input into waveguide 1, and an odd cat is input into waveguide 2, the output is the superposition of two odd cats. Both the cases I ($\alpha_3 > \alpha_1, \alpha_2$) and case III ($\alpha_2 < \alpha_3 < \alpha_1$) will exist in different positions, so we can also amplify the cat by choosing a reasonable coupling length. Let the parameters in Fig. 3 be $\varphi_1 = 0, \varphi_2 = \pi$, so according to Eq. (5), the output is:

$$|\psi_{out}(z)\rangle = \mathcal{N}\left[e^{-\frac{1}{2}(t\beta_0 - r\alpha_0)^2}(|t\alpha_0 + r\beta_0\rangle - |-(t\alpha_0 + r\beta_0)\rangle)\right.$$

$$\left. -e^{-\frac{1}{2}(t\beta_0 + r\alpha_0)^2}(|t\alpha_0 - r\beta_0\rangle - |-(t\alpha_0 - r\beta_0)\rangle)\right] \quad (10)$$

It can be expressed as:

$$|\psi_{out}(z)\rangle = \mathcal{N}[c_1(|\alpha_1\rangle - |-\alpha_1\rangle) + c_2(|\alpha_2\rangle - |-\alpha_2\rangle)]$$

$$\begin{cases} c_1 = e^{-\frac{1}{2}(t\beta_0 - r\alpha_0)^2} \times sign(t\alpha_0 + r\beta_0), c_2 = -e^{-\frac{1}{2}(t\beta_0 + r\alpha_0)^2} \times sign(t\alpha_0 - r\beta_0), \\ \quad \alpha_1 = |t\alpha_0 + r\beta_0|, \alpha_2 = |t\alpha_0 - r\beta_0|; r > 0 \\ c_1 = e^{-\frac{1}{2}(t\beta_0 + r\alpha_0)^2} \times sign(t\alpha_0 - r\beta_0), c_2 = -e^{-\frac{1}{2}(t\beta_0 - r\alpha_0)^2} \times sign(t\alpha_0 + r\beta_0), \\ \quad \alpha_1 = |t\alpha_0 - r\beta_0|, \alpha_2 = |t\alpha_0 + r\beta_0|; r < 0 \end{cases} \quad (11)$$

Here, the signum function appears because $\alpha_1, \alpha_2$ are supposed to be a positive real number in section 2 and $|\alpha_1\rangle - |-\alpha_1\rangle = -(|-\alpha_1\rangle - |\alpha_1\rangle)$). This led to a mutation in the coefficients at the position fulfill $\alpha_0 + r\beta_0 = 0$ or $t\alpha_0 - r\beta_0 = 0$. So, the transition between case I and case III also happened. Let $\alpha_0 = 1.7, \beta_0 = 0.8$ the same as the parameters in Fig. 4, specific results can be obtained, as shown in Fig. 6. A high amplitude cat larger than 2 can be achieved with a high probability in a wide range, similar to the results when two odd cats are input.

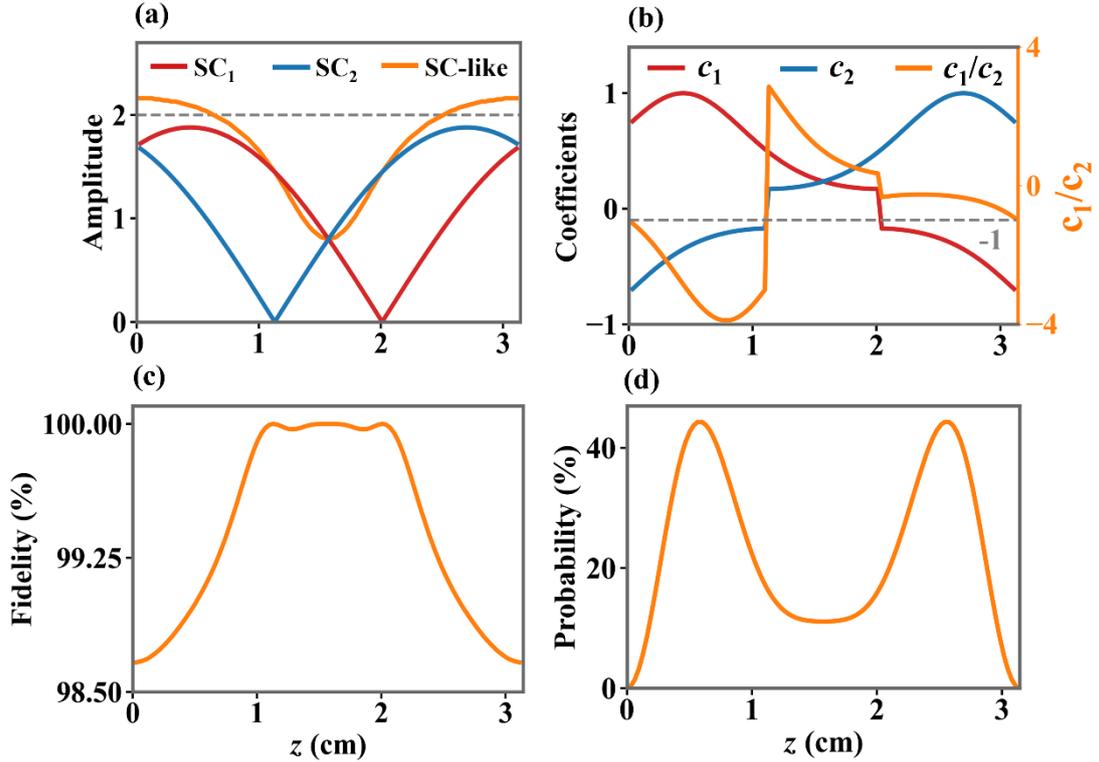

Fig. 6. The characters of SC-like state generated through the superposition of two odd $|SC_1\rangle$ and $|SC_2\rangle$ in coupled waveguide system. Here, the input states are even and odd SCs with $\alpha_0 = 1.7, \beta_0 = 0.8$ respectively. The (a) amplitude and (b) fidelity of the generated SC-like state. (c) The superposition coefficients of $|SC_1\rangle$ and $|SC_2\rangle$ and their ratio. (d) The success probability when zero photons are detected at the output port of waveguide 2.

When an odd cat is input into waveguide 1, and an even cat is input into waveguide 2, the output is also the superposition of two odd cats. Both the cases I ($\alpha_3 > \alpha_1, \alpha_2$) and case III ($\alpha_2 < \alpha_3 < \alpha_1$) will exist. Let the parameters in Fig. 3 be $\varphi_1 = 0, \varphi_2 = \pi$, so according to Eq. (5), the output is:

$$|\psi_{out}(z)\rangle = \mathcal{N}\left[e^{-\frac{1}{2}(t\beta_0 - r\alpha_0)^2}(|t\alpha_0 + r\beta_0\rangle - |-(t\alpha_0 + r\beta_0)\rangle)\right.$$

$$\left.+e^{-\frac{1}{2}(t\beta_0 + r\alpha_0)^2}(|t\alpha_0 - r\beta_0\rangle - |-(t\alpha_0 - r\beta_0)\rangle)\right] \tag{12}$$

It can be expressed as:

$$|\psi_{out}(z)\rangle = \mathcal{N}[c_1(|\alpha_1\rangle - |-\alpha_1\rangle) + c_2(|\alpha_2\rangle - |-\alpha_2\rangle)]$$

$$\begin{cases} c_1 = e^{-\frac{1}{2}(t\beta_0 - r\alpha_0)^2} \times sign(t\alpha_0 + r\beta_0), c_2 = e^{-\frac{1}{2}(t\beta_0 + r\alpha_0)^2} \times sign(t\alpha_0 - r\beta_0), \\ \qquad \alpha_1 = |t\alpha_0 + r\beta_0|, \alpha_2 = |t\alpha_0 - r\beta_0|; r > 0 \\ c_1 = e^{-\frac{1}{2}(t\beta_0 + r\alpha_0)^2} \times sign(t\alpha_0 - r\beta_0), c_2 = e^{-\frac{1}{2}(t\beta_0 - r\alpha_0)^2} \times sign(t\alpha_0 + r\beta_0), \\ \qquad \alpha_1 = |t\alpha_0 - r\beta_0|, \alpha_2 = |t\alpha_0 + r\beta_0|; r < 0 \end{cases} \tag{13}$$

The same as the situation in Fig. 6, the signum function led to the mutation in coefficients and transition between case I and case II. By redesigning the input, such as $\alpha_0 = 0.8, \beta_0 = 1.7$, a high amplitude cat larger than 2 can also be attained, as shown in Fig. 7.

In a word, the cat can be enlarged effectively in three different situations of what we considered by designing the input and coupling length. But there is also a situation that will result in nearly no amplification which should be avoided. These outcomes have great significance in instructing the cat state's breeding by utilizing resources rationally.

## 5. Conclusion

We have revealed that the principle of breeding the cat by superposition of two kittens is that under appropriate coefficients, the two nearby coherent states interfere and become a greater coherent-like state, and then the two coherent-like states superpose and generate a large-size cat-like state. The condition of amplification has also been deduced. Furthermore, we have proposed a scheme to realize the superposition of two cats to generate a large amplitude cat state. Finally, we have demonstrated that we can break through the theoretical limit of previous works with a high success probability. We have considered four situations with different parity of the input cat. Then we found our scheme performs well in three of these situations which indicates that our amplification has a good generality. The principle enhances the understanding of quantum superposition and can be extended to other macroscopic quantum states. Our scheme can be implemented on chips to provide a high-quality cat source that can be used in subsequent quantum information tasks.

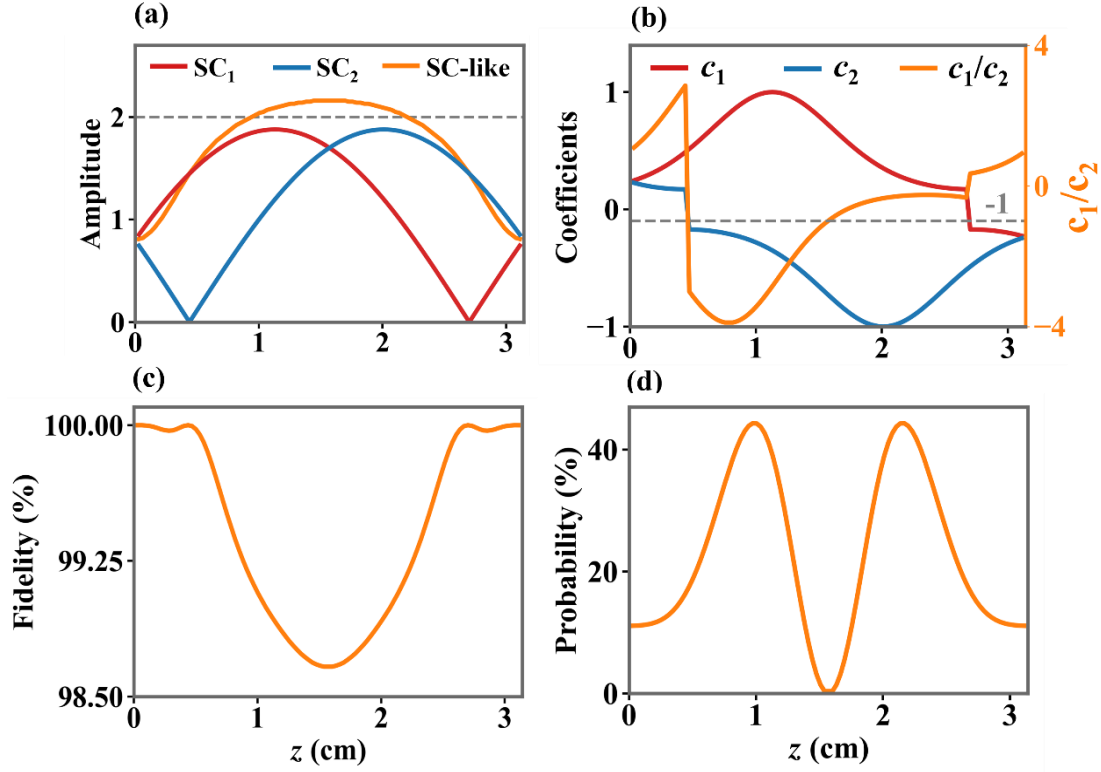

Fig. 7. The characters of SC-like state generated through the superposition of two odd $|SC_1\rangle$ and $|SC_2\rangle$ in coupled waveguide system. Here, the input states are odd and even SCs with $\alpha_0 = 0.8, \beta_0 = 1.7$ respectively. The (a) amplitude and (b) fidelity of the generated SC-like state. (c) The superposition coefficients of $|SC_1\rangle$ and $|SC_2\rangle$ and their ratio. (d) The success probability when zero photons are detected at the output port of waveguide 2.

**Acknowledgments**

This work was supported by the Innovation Program for Quantum Science and Technology (2021ZD0301500) and the National Natural Science Foundation of China (11974032).

**Appendix A. A specific example of the superposition of two coherent states**

In section 2.1, we have deduced the size relation between $\alpha_1, \alpha_2$, and $\alpha_3$ in different cases. Now, a specific example is given here to support these results. We consider the superposition of coherent states $\mathcal{N}(c_1|\alpha_1\rangle + c_2|\alpha_2\rangle)$ with $\alpha_1 = 1.7, \alpha_2 = 1.4, c_1 = 0.2$, and $c_2$ varies from $-0.4$ to $0.2$. Both the Wigner function and fidelity are calculated to enhance persuasiveness.

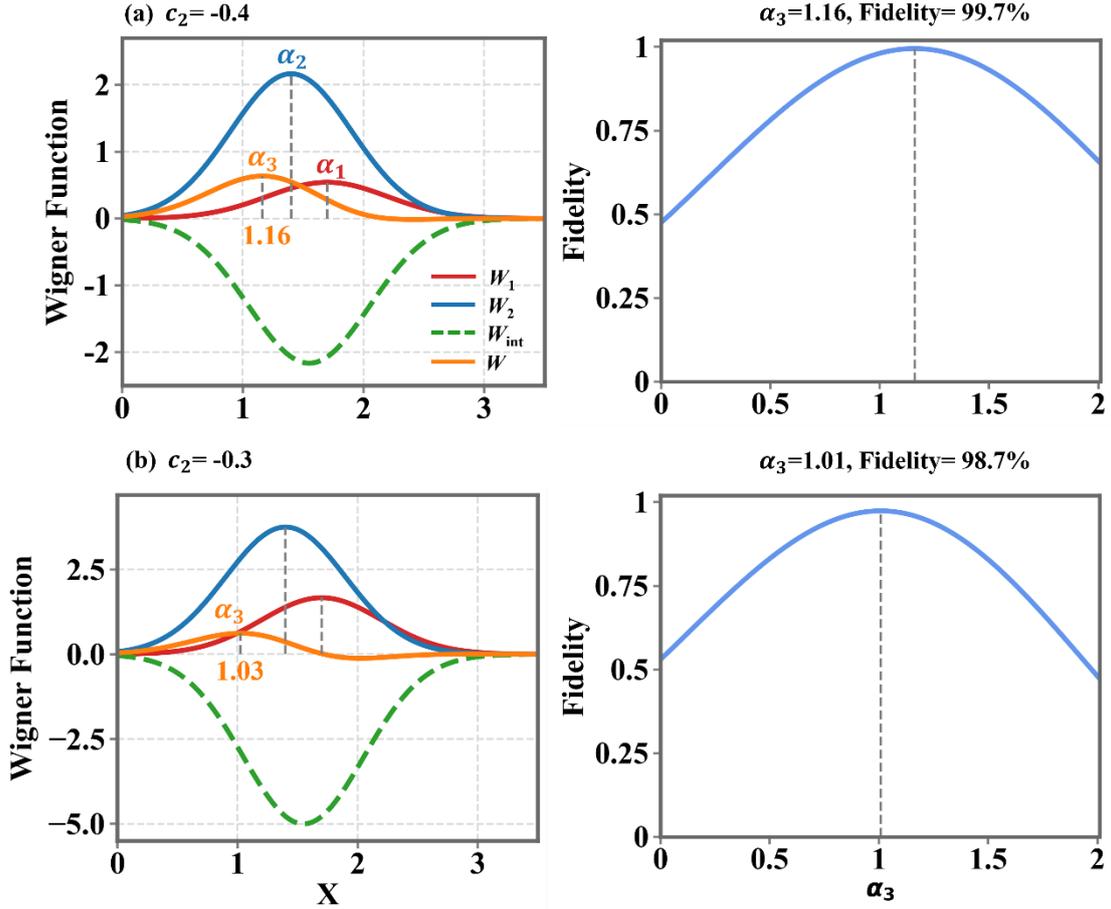

Fig. A1. The Wigner function (at P=0) and the fidelity of the superposed state $\mathcal{N}(c_1|\alpha_1\rangle + c_2|\alpha_2\rangle)$ with $\alpha_1 = 1.7, \alpha_2 = 1.4, c_1 = 0.2$, and $c_1 = -0.4$ (a), and $-0.3$ (b). Both instances support that $\alpha_3 < \alpha_1, \alpha_2$ when $c_1 c_2 < 0, |c_1| < |c_2|$.

① *Case II*: $c_1 c_2 < 0, |c_1| < |c_2|$, here $c_2 = -0.4$, and $-0.3$:

The peak position of the $W(X, 0)$ and the maximal fidelity between the superposed state and coherent state are shown below. There is a slight difference between them in Fig. A1(b) because fidelity cares about the similarity of the whole state, but the peak of the Wigner function only reflects the local property of a state. This deviation would be very small when the fidelity is very high, as shown in Fig. A1(a). However, it's irrelevant to our outcome $\alpha_3 < \alpha_1, \alpha_2$, as shown in Fig. A1. Moreover, the repulsive interference which causes the amplitude to be smaller is strong when the coefficient is closer to $c_2 = -c_1$.

② $c_2 = -c_1 = -0.2$:

The interference is very strong when $c_2 = -c_1$, so the Wigner function of the superposed state $\mathcal{N}(c_1|\alpha_1\rangle + c_2|\alpha_2\rangle)$ can't any longer be similar to a coherent state, and then the superposition of corresponding cat stats behaves in the same way. This situation should be avoided in our scheme if we want to get high-fidelity cat states.

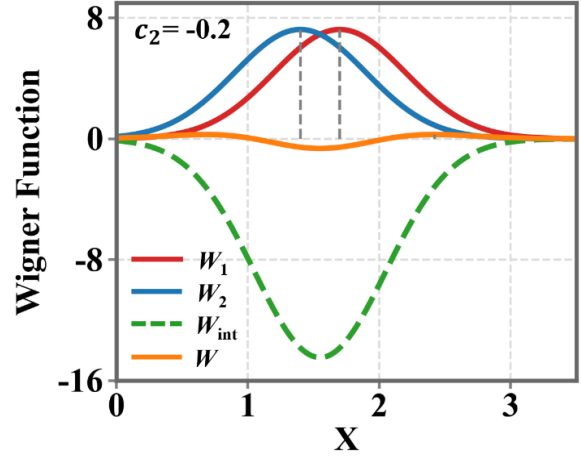

Fig. A2. The Wigner function (at P=0) of the superposed state $\mathcal{N}(c_1|\alpha_1\rangle + c_2|\alpha_2\rangle)$ with $\alpha_1 = 1.7, \alpha_2 = 1.4, c_1 = 0.2$, and $c_1 = -0.2$. The state can't be seen as a coherent-like state because of the strong interference.

③ *Case I*: $c_1c_2 < 0, |c_1| > |c_2|$, here $c_2 = -0.1$, and $-0.05$:

The peak position of the $W(X,0)$ and the maximal fidelity between the superposed state and coherent state are shown in Fig. A3. It's obvious $\alpha_3 > \alpha_1, \alpha_2$ in this situation. The repulsive interference which causes the amplitude to be larger is strong when the coefficient is closer to $c_2 = -c_1$.

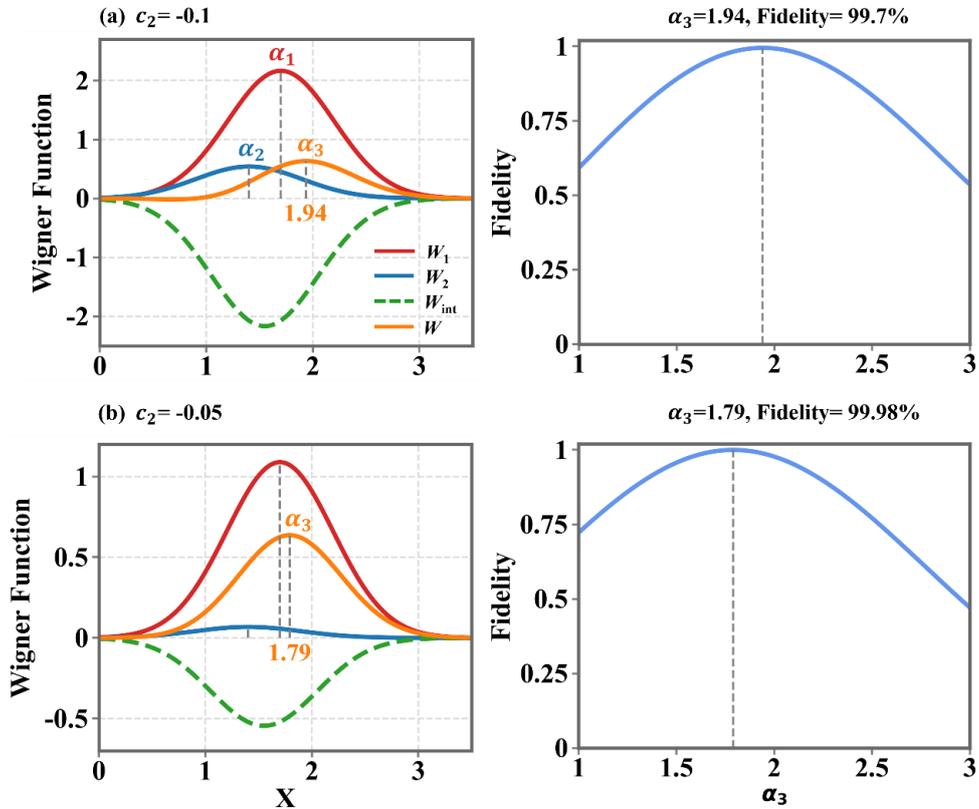

Fig. A3. The Wigner function (at P=0) and the fidelity of the superposed state $\mathcal{N}(c_1|\alpha_1\rangle + c_2|\alpha_2\rangle)$ with $\alpha_1 = 1.7, \alpha_2 = 1.4, c_1 = 0.2$, and $c_1 = -0.1$ (a), and $-0.05$ (b). Both instances support that $\alpha_3 > \alpha_1, \alpha_2$ when $c_1c_2 < 0, |c_1| > |c_2|$.

④ *Case III*: $c_1 c_2 > 0$, here $c_2 = 0.1$, and $0.2$:

The peak position of the $W(X,0)$ and the maximal fidelity between the superposed state and coherent state are shown in Fig. A4. It satisfies $\alpha_2 < \alpha_3 < \alpha_1$ in this situation.

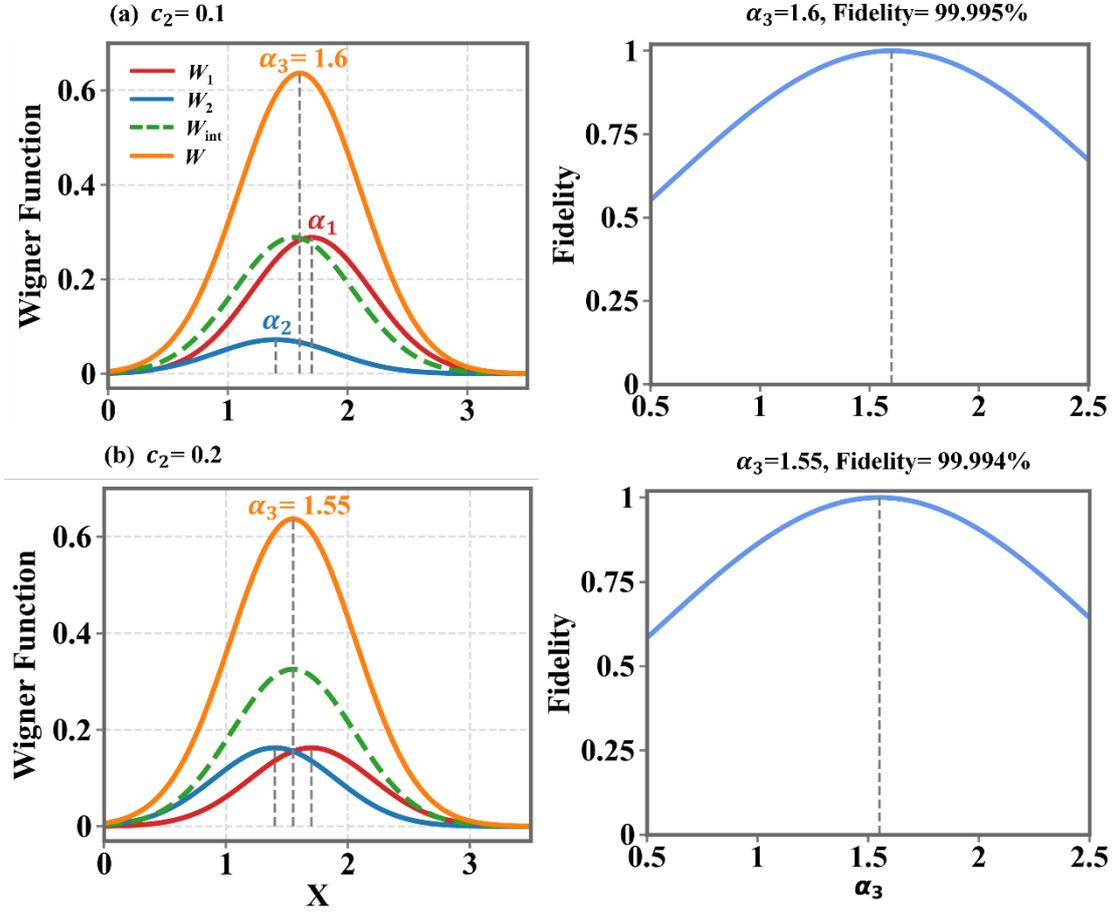

Fig. A4. The Wigner function (at P=0) and the Fidelity of the superposed state $\mathcal{N}(c_1|\alpha_1\rangle + c_2|\alpha_2\rangle)$ with $\alpha_1 = 1.7, \alpha_2 = 1.4, c_1 = 0.2$, and $c_1 = 0.1$ (a), and $0.2$ (b). Both instances support that $\alpha_2 < \alpha_3 < \alpha_1$ when $c_1 c_2 > 0$.

**Appendix B. The output state when detecting m photons in our scheme**

The two-mode output without when two cat states are input into the waveguides is given as Eq. (4). Providing that the coherent state satisfies that

$$|\alpha_c\rangle = e^{-\frac{1}{2}|\alpha_c|^2} \sum_{n=0}^{\infty} \frac{\alpha_c^n}{\sqrt{n!}} |n\rangle \tag{B1}$$

There is

$$|i(t\beta_0 - r\alpha_0)\rangle = e^{-\frac{1}{2}|t\beta_0 - r\alpha_0|^2} \sum_{n=0}^{\infty} \frac{(t\beta_0 - r\alpha_0)^n}{\sqrt{n!}} i^n |n\rangle$$

$$|-i(t\beta_0 - r\alpha_0)\rangle = e^{-\frac{1}{2}|t\beta_0 - r\alpha_0|^2} \sum_{n=0}^{\infty} \frac{(t\beta_0 - r\alpha_0)^n}{\sqrt{n!}} (-i)^n |n\rangle \tag{B2}$$

$$|i(t\beta_0 + r\alpha_0)\rangle = e^{-\frac{1}{2}|t\beta_0 + r\alpha_0|^2} \sum_{n=0}^{\infty} \frac{(t\beta_0 + r\alpha_0)^n}{\sqrt{n!}} i^n |n\rangle$$

$$|-i(t\beta_0 + r\alpha_0)\rangle = e^{-\frac{1}{2}|t\beta_0 + r\alpha_0|^2} \sum_{n=0}^{\infty} \frac{(t\beta_0 + r\alpha_0)^n}{\sqrt{n!}} (-i)^n |n\rangle$$

So the output in waveguide 1 when m photons are measured in waveguide 2 is:

$$|\psi\rangle_m = \mathcal{N} i^m \left\{ e^{-\frac{1}{2}|t\beta_0 - r\alpha_0|^2} \frac{(t\beta_0 - r\alpha_0)^m}{\sqrt{n!}} \left[ |t\alpha_0 + r\beta_0\rangle + (-1)^m e^{i(\varphi_1 + \varphi_2)} |-(t\alpha_0 + r\beta_0)\rangle \right] \right.$$
$$\left. + e^{-\frac{1}{2}|t\beta_0 + r\alpha_0|^2} \frac{(t\beta_0 + r\alpha_0)^m}{\sqrt{n!}} \left[ (-1)^m \times e^{i\varphi_2} |t\alpha_0 - r\beta_0\rangle + e^{i\varphi_1} |-(t\alpha_0 - r\beta_0)\rangle \right] \right\} \tag{B3}$$

It's the superposition of two cat states with an amplitude equal to $|t\alpha_0 + r\beta_0|$ and $|t\alpha_0 - r\beta_0|$ respectively. The Eq. (5) can be easily attained by letting $m = 0$ in Eq. (B3).